\documentclass[fleqn,10pt]{wlscirep}
\usepackage{graphicx}
\usepackage{subcaption}
\title{Effective Three-Body Interactions in Jaynes-Cummings-Hubbard Systems}

\author[1,*]{Srivatsa B. Prasad}
\author[1]{Andrew M. Martin}
\affil[1]{School of Physics, The University of Melbourne, Parkville, 3010, Australia}

\affil[*]{srivatsa.badariprasad@unimelb.edu.au}

\begin{abstract}
A generalisation of the Jaynes-Cummings-Hubbard model for coupled-cavity arrays is introduced, where the embedded two-level system in each cavity is replaced by a $\Xi$-type three-level system. We demonstrate that the resulting effective polariton-polariton  interactions at each site are both two-body and three-body. By tuning the ratio of the two transition dipole matrix elements, we show that the strength and sign of the two-body interaction can be controlled whilst maintaining a three-body repulsion. We then proceed to demonstrate how different two-body and three-body interactions alter the mean field superfluid-Mott insulator phase diagram, with the possible emergence of a pair superfluid phase in the two-body attractive regime.
\end{abstract}
\begin{document}

\flushbottom
\maketitle

\thispagestyle{empty}

\section*{Introduction}
Recently, there has been considerable effort applied to the study of solid-state phenomena in Jaynes-Cummings-Hubbard (JCH) systems. These studies have predicted a superfluid-Mott insulator phase transition \cite{natphys_2_12_849-855_2006,natphys_2_12_856-861_2006,pra_76_3_031805_2007,prl_103_8_086403_2009,pra_80_2_023811_2009,prl_104_21_216402_2010,pra_81_6_061801r_2010}, supersolid \cite{pra_90_4_043801_2014} and Bose-glass \cite{prl_99_18_186401_2007} phases, metamaterial properties \cite{optexpress_19_12_11018-11033_2011}, and, in the presence of time-reversal-symmetry breaking, fractional quantum Hall states \cite{prl_101_24_246809_2008,prl_108_20_206809_2012,prl_108_22_223602_2012,pra_93_5_053614_2016}. Almost all of these studies have focused on the case where each lattice element of the system is comprised of a two-level system interacting with a single photonic mode, resulting in the emergence of quasiparticles, referred to as polaritons. 

In this work we consider the properties of a generalised JCH system where each lattice element is comprised of a three-level system interacting with a single photon mode. Such a model provides an opportunity to construct a Hamiltonian in which three-body interactions are dominant. The effects of such interactions have been of great interest for some time, in part due to their connection to the Pfaffian state that plays a key role in the theory of the fractional quantum Hall effect \cite{nuclphysb_360_2-3_362-396_1991}. For example, with a view to better understanding these Pfaffian states, methods of introducing effective three-body interactions in dilute gas Bose-Einstein condensates have been proposed \cite{natphys_3_10_726-731_2007,prl_102_4_040402_2009,pra_82_4_043629_2010,pra_89_5_053619_2014}. In the context of the Bose-Hubbard model \cite{prb_40_1_546-570_1989}, it has been predicted that three-body interactions can significantly change the superfluid-Mott insulator phase diagram enabling a pair superfluid phase to manifest itself \cite{prl_102_4_040402_2009,prb_82_6_064509_2010,prb_82_6_064510_2010,pra_81_6_061604_2010,prl_106_18_183302_2011,pra_92_4_043615_2015}, for attractive (repulsive) two-body (three-body) interactions. Accordingly, here, we consider a three-level JCH system which supports both two-body and three-body effective contact interactions, on each site of the JCH lattice, and analyse the zero-temperature phases that this system exhibits. 

\section*{Results}
\begin{figure}[ht]
\centering
\includegraphics[width=0.35\linewidth]{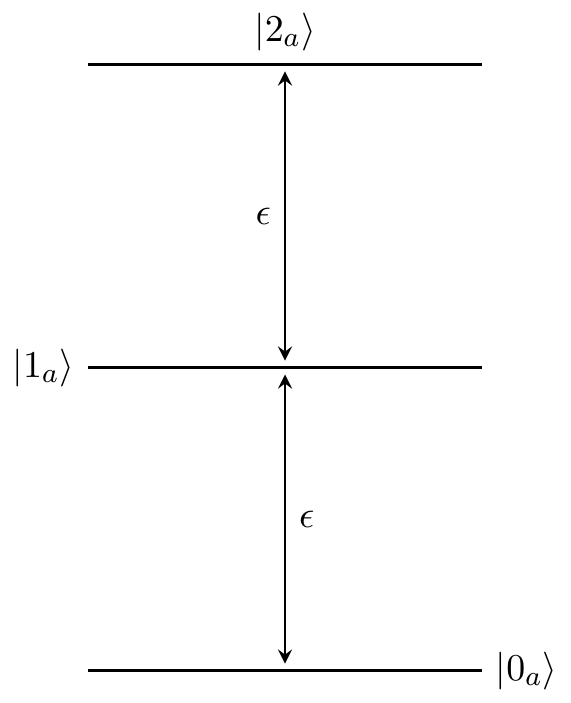}
\caption{The $\Xi$-type (cascade) configuration for a dipole-allowed three-level system, with an identical level spacing $\epsilon$.}
\label{tikztopdf}
\end{figure}
\subsection*{Three-Level Jaynes-Cummings Cavity}
In the standard two-level coupled atom cavity Jaynes-Cummings system \cite{procieee_51_1_89-109_1963} an effective two-body repulsion, for the polariton quasiparticles, is provided by the interactions between the two-level atom and the photons. In this work we wish to consider the impact of three-body interactions and the interplay between two-body and three-body interactions on the properties of the mean-field superfluid-Mott insulator phase diagram. As such we  introduce an additional, local, three-body interaction. To achieve this, we generalise the two-level system on each site to a three-level system. There exist a number of possible configurations for a cavity containing a three-level system. Parity considerations restrict the number of viable transition configurations to three - the cascade ($\Xi$), $\Lambda$ and $V$-type \cite{klimovchumakovqtmoptics}. However, the restriction to a single cavity mode and the use of the rotating wave approximation \cite{klimovchumakovqtmoptics,atomicphysicsfoot} necessitates the use of solely the cascade configuration, represented in Fig. \ref{tikztopdf}.

\subsubsection*{The Three-Level Cascaded Jaynes-Cummings Cavity}
For simplicity, we model an ideal system with three atomic energy levels equally spaced by $\epsilon$, see Fig.~\ref{tikztopdf}, and a resonant cavity mode with associated energy $\omega$. This results in a coupled-cavity array identical to one that has previously been proposed as a platform for the emulation of Pfaffian states via cavity QED \cite{pra_93_5_053614_2016}. As we assume that the three-level system is in the $\Xi$-configuration, we allow for photon-mediated atomic transitions only between the $|0_a\rangle \rightarrow |1_a \rangle$ and $|1_a\rangle \rightarrow  |2_a \rangle$ atomic levels; we denote the corresponding transition dipole moments as $\beta_{01}$ and $\beta_{12}$ respectively, with $\beta_{02}=0$. We also introduce the photonic creation and annihilation operators ${\hat a}^{\dagger}$ and ${\hat a}$, along with the atomic operators $\lambda^+_{\downarrow}=\left(\lambda^-_{\downarrow}\right)^{\dagger}$ and  $\lambda^+_{\uparrow}=\left(\lambda^-_{\uparrow}\right)^{\dagger}$, where $\lambda^+_{\downarrow}$ and $\lambda^+_{\uparrow}$ excite the atom from $|0_a\rangle \rightarrow |1_a \rangle$ and $|1_a\rangle \rightarrow  |2_a \rangle$ respectively. Subsequently, with $\widetilde{\beta} = \beta_{12}/\beta_{01}$, the application of the rotating wave approximation results in the Hamiltonian for a single site taking the form:  
\begin{equation}
\widehat{H}^{JC3} = -\widetilde{\mu}'\hat{n} - \widetilde{\Delta}\hat{n}_{a} + \widetilde{\beta}\left(\hat{\lambda}^{+}_{\uparrow}\hat{a}+\hat{\lambda}^{-}_{\uparrow}\hat{a}^{\dagger}\right) + \hat{\lambda}^{+}_{\downarrow}\hat{a}+\hat{\lambda}^{-}_{\downarrow}\hat{a}^{\dagger}. \label{eq:jc3}
\end{equation}
Here $\hat{n} = \hat{n}_{ph} + \hat{n}_{a}$ characterises the total number of excitations in the system, with $\hat{n}_{ph} = \hat{a}^{\dagger}\hat{a}$ and $\hat{n}_{a} = 2\hat{\lambda}^{+}_{\uparrow}\hat{\lambda}^{-}_{\uparrow}+\hat{\lambda}^{+}_{\downarrow}\hat{\lambda}^{-}_{\downarrow}$ corresponding to the number of photonic and atomic excitations respectively. The parameters $\widetilde{\mu}' = (\mu - \omega)/\beta_{01}$ and $\widetilde{\Delta} = (\omega - \epsilon)/\beta_{01}$ represent the rescaled chemical potential and cavity detuning respectively. Since $\hat{n}$ commutes with $\widehat{H}^{JC3}$, the Hamiltonian for this system may be diagonalised by decomposing it into a direct sum of operators $\widehat{H}_n^{JC3}$, each acting on the Fock subspace spanned by basis states $|n_a,n_{ph}\rangle\equiv|n_a\rangle\otimes|n_{ph}\rangle$ where $n_a+n_{ph}=n$.

For each $n$-particle subspace, we work in the ordered basis $\left\lbrace|0,n\rangle,\cdots,|i,n-i\rangle\right\rbrace$, where $i = 0$ for $n = 0$, $i = 1$ for $n = 1$ and $i = 2$ for $n\geq 2$. For the vacuum, where $n = 0$, $\widehat{H}_0^{JC3}$ is trivially zero. For the one-particle subspace ($n = 1$), $\hat{\lambda}^{\pm}_{\uparrow}$ is undefined and the usual Pauli matrix-based representation applies to the $\hat{\lambda}^{\pm}_{\downarrow}$ operators \cite{procieee_51_1_89-109_1963}. However, for $n \geq 2$, these new atomic operators are representable via the \textit{Gell-Mann matrices}, generators of su$(3)$\cite{mathmethphysarfkenweber}, given by
\begin{eqnarray}
\hat{\lambda}^{+}_{\uparrow} &= \left(\hat{\lambda}^{-}_{\uparrow}\right)^{\dagger} = \frac{1}{2}\left(\hat{\lambda}_1 + i\hat{\lambda}_2\right) = 
\left(\begin{array}{ccc}
0 & 1 & 0 \\
0 & 0 & 0 \\
0 & 0 & 0
\end{array}\right), \,\,\,
\hat{\lambda}^{+}_{\downarrow} &= \left(\hat{\lambda}^{-}_{\downarrow}\right)^{\dagger} = \frac{1}{2}\left(\hat{\lambda}_6 + i\hat{\lambda}_7\right) = 
\left(\begin{array}{ccc}
0 & 0 & 0 \\
0 & 0 & 1 \\
0 & 0 & 0
\end{array}\right). \nonumber
\end{eqnarray}
This yields the matrix blocks to be diagonalised for each $n$, given by
\begin{eqnarray}
\widehat{H}_0^{JC3} = \left(\begin{array}{c} 0 \end{array}\right), \,\,\,
\widehat{H}_{1}^{JC3} = -\widetilde{\mu}'\hat{\mathbb{I}} + \left(\begin{array}{cc}
-\widetilde{\Delta} & 1 \\
1 & 0
\end{array}\right), \,\,\,
\widehat{H}_{n\geq 2}^{JC3} = -n\widetilde{\mu}'\hat{\mathbb{I}} + 
\left(\begin{array}{ccc}
-2\widetilde{\Delta} & \widetilde{\beta}\sqrt{n-1} & 0 \\
\widetilde{\beta}\sqrt{n-1} & -\widetilde{\Delta} & \sqrt{n} \\
0 & \sqrt{n} & 0
\end{array}\right). \label{eq:jc3_2}
\end{eqnarray}

We initially detail the nature of the spectrum of Eq.~(\ref{eq:jc3}). The spectra for $n = 0,1$ are necessarily identical to those in the Jaynes-Cummings model \cite{klimovchumakovqtmoptics} with the existence of two distinct eigenstates for $n = 1$. However, it is readily seen from Eq.~(\ref{eq:jc3_2}) that there must exist three independent eigenstates for $n > 2$. The energies of each of these branches is \cite{mathgaz_77_480_354-359_1993,physscr_76_3_244-248_2007} given by
\begin{eqnarray}
E(n = 0) &=& 0, \label{eq:En0} \\
E_{\pm}(n = 1) &=& -\widetilde{\mu}' - \frac{\widetilde{\Delta}}{2} \pm \sqrt{1 + \frac{\widetilde{\Delta}^2}{4}} , \label{eq:En1}\\
E_k(n\geq 2) &=& -n\tilde{\mu}' - \widetilde{\Delta} + \xi_k^{(n)}\,;\,k\in\left\lbrace 1,2,3\right\rbrace, \label{eq:En2}
\end{eqnarray}
where
\begin{eqnarray}
\xi_k^{(n)} &=& u_n\cos\left(\theta_n + \frac{2\pi k}{3}\right)\,;\,k\in\left\lbrace 1,2,3\right\rbrace, \label{eq:16}\\
 \theta_n &=& \frac{1}{3}\arccos\left(\frac{-4q_n}{u_n^3}\right),\label{eq:arcosine} \\
u_n &=& 2\sqrt{\frac{p_n}{3}}, \,\,\, 
p_n = \widetilde{\Delta}^2 + (n-1)\widetilde{\beta}^2 + n \,\,\, {\rm and} \,\,\,
q_n = \widetilde{\Delta}\left((n-1)\widetilde{\beta}^2 - n\right). \label{eq:coeffs_xi}
\end{eqnarray}
The Fock state representations for the corresponding \textit{polariton} branches are provided by \cite{mathgaz_77_480_354-359_1993}
\begin{eqnarray}
| n = 0 \rangle &=& | 0,0\rangle, \label{eq:unpert_1}\\
|n = 1\rangle_{\pm} &=& \frac{|0,1\rangle + \left(-\frac{\widetilde{\Delta}}{2}\pm\sqrt{1 + \frac{\widetilde{\Delta}^2}{4}}\right)|1,0\rangle}{\sqrt{2\left(1 + \frac{\widetilde{\Delta}^2}{4}\right)\mp\widetilde{\Delta}\sqrt{1 + \frac{\widetilde{\Delta}^2}{4}}}}, \label{eq:unpert_2}\\
|n\geq 2\rangle_k &=& \frac{\left[\xi_k^{(n)}\left(\xi_k-\widetilde{\Delta}\right)-n\right]|2,n-2\rangle}{\sqrt{\zeta_k^{(n)}}} + \frac{\sqrt{n-1}\left[\widetilde{\beta}\left(\xi_k^{(n)}-\widetilde{\Delta}\right)|1,n-1\rangle + \sqrt{n}\widetilde{\beta}|0,n\rangle\right]}{\sqrt{\zeta_k^{(n)}}}  \label{eq:unpert_3},
\end{eqnarray}
where for $n\ge2$ the normalisation factor $\zeta_k^{(n)}$ is defined as
\begin{eqnarray}
\zeta_k^{(n)}&=&\left[\xi_k^{(n)}\left(\xi_k^{(n)}-\widetilde{\Delta}\right)-n\right]^2+(n-1)\widetilde{\beta}^2\left(\xi_k^{(n)}-\widetilde{\Delta}\right)^2 +n(n-1)\widetilde{\beta}^2.
\end{eqnarray}
It is necessary to be able to identify which branch ($k$) has the lowest energy for a given polariton number $n$. If the arccosine in Eq.~(\ref{eq:arcosine}) is restricted to its principle branch, i.e. $[0,\pi]$, then $\theta_n \in \left[0,\frac{\pi}{3}\right]\,\forall\,n\geq 2$ and thus
\begin{equation}
E_1(n) < E_2(n) < E_3(n).
\end{equation}
Therefore, the ground state for any polariton number $n\geq 2$ may be taken to be the $k = 1$ branch, or equivalently the `$-$' branch for $n = 1$, and for future reference we denote the dimensionless energies corresponding to these branches as $E_g(n)$.

\begin{figure}
\centering
\includegraphics[width=\linewidth]{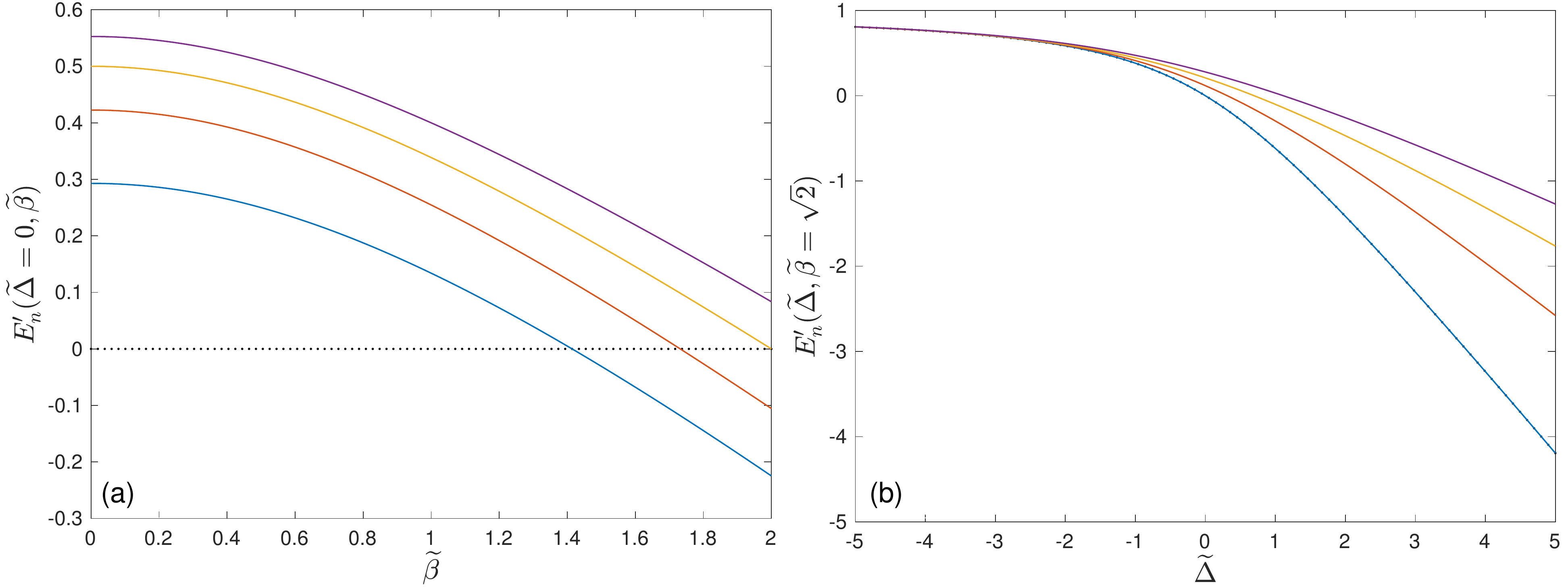}
\caption{The rescaled polariton energy, $E_n'=E_g(n)/n$ for $\widetilde{\mu}' = -1$: (a) as a function of $\widetilde{\beta}$ for $\widetilde{\Delta} = 0$; (b) as a function of $\widetilde{\Delta}$ for $\widetilde{\beta} = \sqrt{2}$. In both (a) and (b) the dotted curve is for $n = 1$ while the solid blue, brown, yellow and purple curves are for $n = 2$, $3$, $4$ and $5$ respectively.}
\label{energyvsdeltaandbeta}
\end{figure}

An effective two-body repulsion for photons is provided in a Jaynes-Cummings cavity by the interactions between the two-level atom and the photons; this is the cause of the photon blockade effect \cite{pra_46_11_r6801r_1992,nature_436_7047_87-90_2005}. In the three-level system coupled to a single cavity mode, the existence of the third atomic level, and its interaction with the cavity mode, induces an additional three-body nonlinearity \cite{pra_93_5_053614_2016}. Furthermore, in this case, the sign of neither the two-body or three-body nonlinearities are immediately evident, but may nonetheless be discerned from inspecting the eigenenergies of the lowest energy polariton branches for each occupation $n$. For each $n$, we initially define the quantity $E_g(n)$ to be the eigenenergy of the respective lowest polariton branch, i.e. $E_1(n)\,\forall\,n\geq 2$, $E_{-}(1)$ for $n = 1$ and $E(0)$ for $n = 0$. The quantity $E_n' = E_g(n)/n$ is thus the energy of each polariton quasiparticle in the ground state corresponding to an occupation of $n$ polaritons in a cavity. Let us initially consider the case where there is a single polariton in the cavity. If the two-body nonlinearity is repulsive, there must be an energy cost associated with adding a second polariton to the cavity, i.e. $E_2' > E_1'$. Conversely, a two-body attraction is associated with an energy cost associated with removing a polariton from a cavity with two polaritons, a scenario marked by the property $E_2' < E_1'$. We thus plot $E_n'$ as a function of $\widetilde{\beta}$, at $\widetilde{\Delta} = 0$, in Fig. \ref{energyvsdeltaandbeta}(a), in order to determine the sign of the two-body nonlinearity. From this it is evident that as $\widetilde{\beta}$ is increased, the proportional energies for one (the dotted curve) and two (the blue curve) polaritons cross, suggesting that a crossover of the two-body interaction, from repulsion to attraction, occurs at a critical value of $\widetilde{\beta}$. As for the three-body nonlinearity, it is readily seen that $E_2' < E_3'$, and indeed $E_n' < E_{n+1}'\,\forall\,n\geq 2$, for all $\widetilde{\beta}$. Since there is always an energy cost associated with adding an additional polariton to a cavity with two polaritons already present, we conclude that the three-body nonlinearity is not only repulsive, but sufficiently so to ensure the presence of the energy cost even when the two-body nonlinearity is an attraction.

In the Methods section, we show that taking the difference of the proportional energy $E_n'$ for $n = 1,2$ yields the critical value $\widetilde{\beta}^2 = 2$, for which $E_1' = E_2'$, which is independent of the detuning $\widetilde{\Delta}$. Note that this result may also be obtained from a mapping onto a two-mode bosonic system, with intermode tunnelling and a three-body constraint \cite{pra_93_5_053614_2016}. That the two-body repulsion-attraction crossover is independent of $\widetilde{\Delta}$ is numerically demonstrated in Fig. \ref{energyvsdeltaandbeta}(b), in which the $E_n'$ is plotted as a function of $\widetilde{\Delta}$ at $\widetilde{\beta}^2 = 2$. In this regime, the energy per polariton is degenerate for $n = 1$ and $n = 2$ regardless of $\widetilde{\Delta}$, which is evident in the coincidence of the blue and dotted lines in Fig. \ref{energyvsdeltaandbeta}(b). This figure also demonstrates that the three-body nonlinearity is a repulsion, irrespective of $\widetilde{\beta}$ and $\widetilde{\Delta}$.

\subsection*{Jaynes-Cummings-Hubbard Mean-Field Phase Diagram}
With the exact single-site solutions in hand we now consider a lattice of such systems with photon hopping permissible between nearest neighbours. For such a system, we wish to  compute the resulting zero-temperature mean-field phase diagram. A reasonable first guess would be that the superfluid and Mott insulator phases also exist in this modified system, and thus we proceed to apply the mean-field approximation to the hopping terms to yield a purely local Hamiltonian. Explicitly, this involves the substitution of the expression \cite{pra_63_5_053601_2001}
\begin{eqnarray}
\sum_{\left\langle i,j\right\rangle}\hat{a}_i^{\dagger}\hat{a}_j &\approx& z\psi\sum_i\left(\hat{a}_i^{\dagger}+\hat{a}_i\right) - z\psi^2, \\
\psi &=& \langle\hat{a}_i\rangle\,\forall\,i,
\end{eqnarray}
where $z$ is the number of nearest-neighbour sites, a quantity that is dependent on the lattice geometry (the superfluid order parameter $\psi$ is assumed to be real). This yields an energy approximation consisting of a sum over all sites of the Hamiltonian
\begin{equation}
\widehat{H}^{JCH3}_{MF} = \widehat{H}^{JC3} - \widetilde{\kappa}\psi\left(\hat{a}^{\dagger}+\hat{a}\right) + \widetilde{\kappa}\psi^2 \label{eq:jch3_MF}
\end{equation}
where $\widetilde{\kappa} = z\kappa/\beta_{01}$.

\begin{figure*}[ht]
\centering
\includegraphics[width=0.9\linewidth]{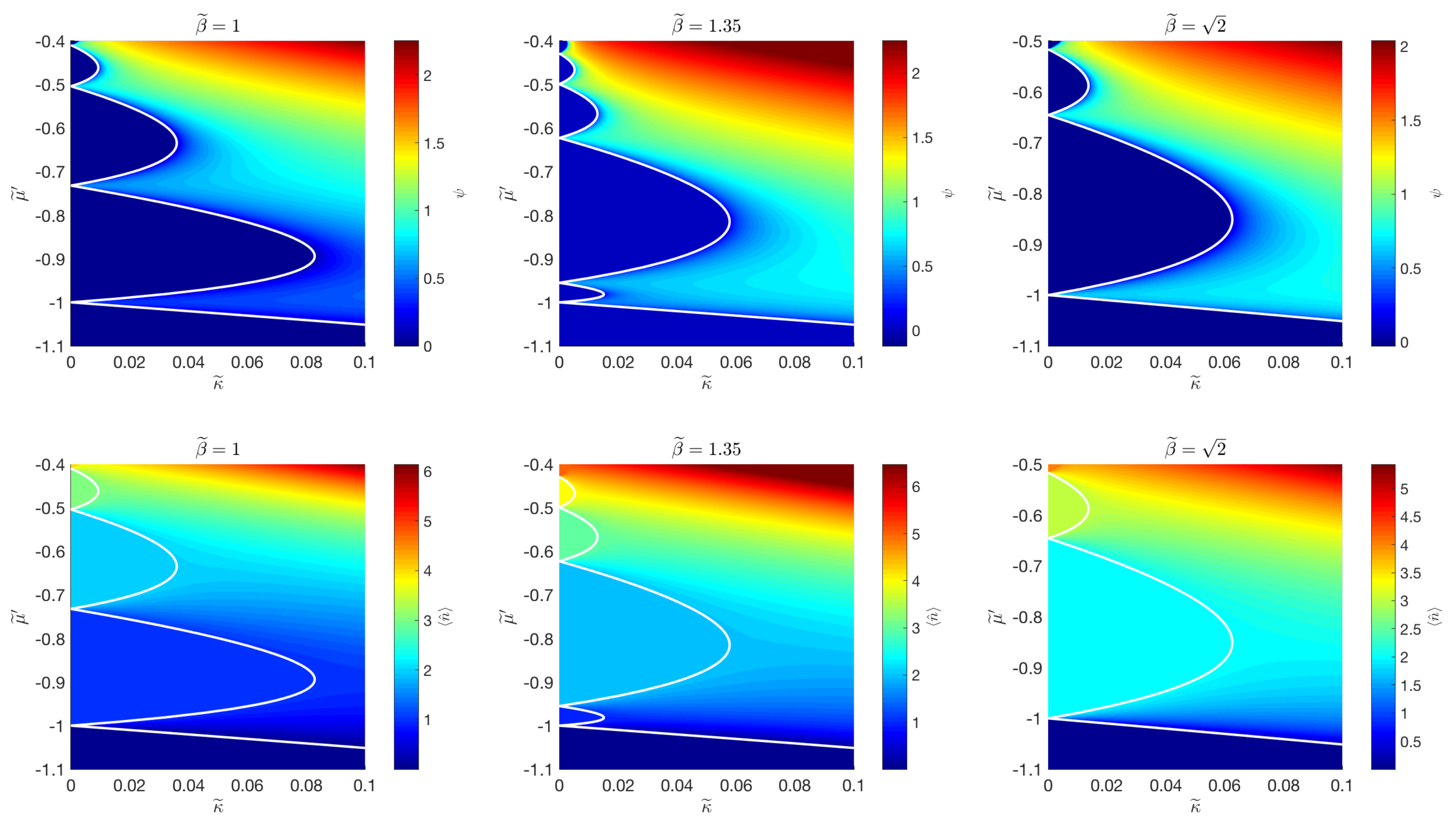}
\caption{Phase and excitation diagrams for $\widetilde{H}_{MF}^{JCH3}$ at $\widetilde{\beta} = \left\lbrace 1,1.35,\sqrt{2}\right\rbrace$. The top row plots the superfluid order parameter, $\psi$, as a function of $\widetilde{\kappa}$ and $\widetilde{\mu}'$, for $\widetilde{\Delta} = 0$. The bottom row features corresponding plots of the expectation value of the polariton number, $\langle\hat{n}\rangle$. The white lines are superfluid-Mott insulator phase boundaries as determined by perturbation theory.}
\label{phasenumbercomparison}
\end{figure*}

To numerically diagonalise the Hamiltonian in Eq.~(\ref{eq:jch3_MF}), we employ a \textit{self-consistent field} approach, which we shall briefly detail here. Initially, we truncate the single-site basis to Fock states $|n_a,n_{ph}\rangle$ such that $n_a + n_{ph} \leq N_{\text{max}}$, where, in this work, we specify the maximum polariton occupation to be $N_{\text{max}} = 16$. With an initial guess for $\psi$, we diagonalise $\widehat{H}^{JCH3}_{MF}$ with respect to this truncated basis, and check that the resulting ground state respects $\langle\hat{a}\rangle = \psi$. If it does not, the obtained $\langle\hat{a}\rangle$ is used as a new guess for $\psi$, and the process is repeatedly iterated until adequate convergence is achieved.

Subsequently, the phase diagram may be obtained from the solution space by plotting the self-consistently determined values of $\psi$ in a given region of parameter space. However, analytical expressions for the phase boundaries may also be found via Landau theory, provided that the transitions are second-order continuous. Following Landau's prescription, the free energy (which reduces to the many-body ground state energy in the zero temperature limit) may be expressed as a power series in the order parameter $\psi$, i.e. \cite{pra_63_5_053601_2001, pra_80_2_023811_2009}
\begin{equation}
E = a_0 + a_2\vert\psi\vert^2 + a_4\vert\psi\vert^4 + \cdots \label{eq:landau}
\end{equation}
At a continuous phase transition, minimisation of the free energy with respect to $\psi$ yields the criterion that the phase boundary is demarcated by solutions to the curve \cite{sachdevquantumpt}$a_2 = 0$. Solutions to this are especially simple to find if the order parameter $\psi$ is manifest in a perturbative expansion of the ground state energy as a power series, which one finds to be the case in Hubbard-type systems \cite{pra_63_5_053601_2001}. To construct this expansion, we note that Eq.~(\ref{eq:jch3_MF}) is purely diagonal except for the term $\widetilde{\kappa}\psi\left(\hat{a}^{\dagger}+\hat{a}\right)$, and so it is partitioned as
\begin{equation}
\widehat{H}^{JCH3}_{MF} = \widehat{H}_0 + \psi\widehat{V}, \label{eq:ptsplit}
\end{equation}
where
\begin{equation}
\widehat{H}_0 = -\widetilde{\mu}'\hat{n} - \widetilde{\Delta}\hat{n}_{a} + \left(\widetilde{\beta}\hat{\lambda}^{+}_{\uparrow}+\hat{\lambda}^{+}_{\downarrow}\right)\hat{a} + \left(\widetilde{\beta}\hat{\lambda}^{-}_{\uparrow}+\hat{\lambda}^{-}_{\downarrow}\right)\hat{a}^{\dagger} + \widetilde{\kappa}\psi^2, \label{eq:ptexact}
\end{equation}
and $\widehat{V} = \widetilde{\kappa}\psi\left(\hat{a}^{\dagger}+\hat{a}\right)$.
Then, we can employ Rayleigh-Schrodinger perturbation theory to expand each perturbed polariton eigenstate's energy in a power series of $\widetilde{\kappa}\psi$ of the form\cite{sakuraimodernqm}
\begin{eqnarray}
E_g &=& E_g^{(0)} + \widetilde{\kappa}\psi E_g^{(1)} + \widetilde{\kappa}^2\psi^2 E_g^{(2)} + \cdots, \label{eq:pert_expansion}\\
E_g^{(0)} &=& \langle g |\widehat{H}_0 |g\rangle, \label{eq:Eg_0}\\
E_g^{(1)} &=& \langle g |\widehat{V} |g \rangle, \label{eq:Eg_1}\\
E_g^{(2)} &=& \sum_{m\neq g}\frac{\vert\langle g| \widehat{V} |m\rangle \vert^2}{E_g^{(0)}-E_m^{(0)}} \label{eq:pert_second}.
\end{eqnarray}
For the unperturbed reference eigenstates Eqs.~(\ref{eq:unpert_1},\ref{eq:unpert_2},\ref{eq:unpert_3}), the quantity $\langle n\rangle$ is clearly a good quantum number, implying that, by Eq.~(\ref{eq:Eg_1}), the first-order correction $E_g^{(1)}$ vanishes identically. Subsequently, a comparison of Eqs.~(\ref{eq:landau}) and (\ref{eq:pert_expansion}) suggests that we make the identification that \cite{pra_80_2_023811_2009}
\begin{eqnarray}
E_g^{(0)} &=& a_0 + \widetilde{\kappa}\psi^2 \label{eq:landau_0}\\
E_g^{(2)} &=& a_2 - \widetilde{\kappa}\psi^2. \label{eq:landau_2}
\end{eqnarray}
The calculation of the second-order perturbative correction is somewhat tedious; we emphasise that in the summation over accessible virtual states in Eq.~(\ref{eq:pert_second}), the contributions from every polariton branch must be considered and not merely the lowest-energy branches for a given polariton occupation. For the sake of clarity, the full expressions for $E_n^{(2)}$ can be found in the Methods section. Below, we merely compare the predictions for the phase boundary obtained via the numerical self-consistent field and analytical Landau-perturbative methods. For simplicity, the analysis is limited to the system with zero detuning, that is, $\widetilde{\Delta} = 0$. Nontheless, it is possible to extend these results to a nonzero detuning in the analytical method, as well as in the numerical formalism. 

\subsubsection*{Two-Body Repulsion}
We first investigate the regime in which the two-body interaction is predicted to be repulsive, i.e. $\widetilde{\beta} < \sqrt{2}$. In Fig. \ref{phasenumbercomparison}, the superfluid-to-Mott insulator phase diagram is presented at $\widetilde{\Delta} = 0$ for the regimes $\widetilde{\beta} = \left\lbrace 1,1.35,\sqrt{2}\right\rbrace$. It is clear that for $\widetilde{\beta} = 1$ (the first column in Fig. \ref{phasenumbercomparison}), the phase diagram closely resembles that of the two-level JCH system, i.e. successive Mott lobes corresponding to successive integer $\langle \hat{n}\rangle$. However, as $\widetilde{\beta}$ is increased to $1.35$ (the middle column in Fig. \ref{phasenumbercomparison}), the first Mott lobe ($\langle \hat{n}\rangle = 1$) shrinks. At $\widetilde{\beta} = \sqrt{2}$ the $\langle \hat{n}\rangle = 1$ Mott lobe vanishes, with the $\langle \hat{n}\rangle = 2$ Mott lobe being adjacent to the vacuum ($\langle\hat{n}\rangle = 0$).

\subsubsection*{Two-Body Attraction}
\begin{figure}[ht]
\centering
\includegraphics[width=\linewidth]{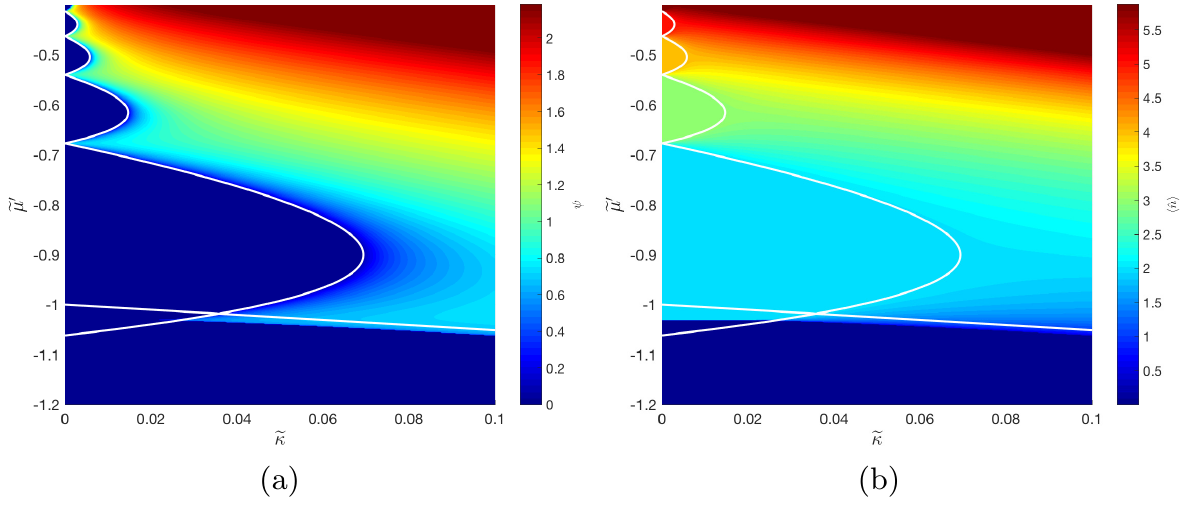}
\caption{Superfluid order parameter $\psi$, (a), and polariton number expectation value $\langle\hat{n}\rangle$, (b), for $\widehat{H}^{JCH3}_{MF}$  as functions of $\widetilde{\kappa}$ and $\widetilde{\mu}'$ for $\widetilde{\beta} = 1.5$ and $\widetilde{\Delta} = 0$. The white lines are the superfluid-Mott insulator phase boundaries as determined by perturbation theory.}
\label{jch3_betaequals1point5_psi_and_n}
\end{figure}

We now focus on the regime in which we predict the existence of an attractive two-body interaction, that is, when $\widetilde{\beta} \geq 2$. As $\widetilde{\beta}\rightarrow\sqrt{2}$ from below, it was found, see Fig. \ref{phasenumbercomparison}, that the first Mott lobe shrank and subsequently vanished at the threshold of $\widetilde{\beta} = \sqrt{2}$, raising the question of what features might be seen in the attractive regime. Retaining $\widetilde{\Delta} = 0$, we cross the threshold and examine the numerical and analytical predictions for the phase diagram at $\widetilde{\beta} = 1.5$ in Fig. \ref{jch3_betaequals1point5_psi_and_n}(a). The analytical prediction is a first Mott lobe in the negative $\widetilde{\kappa}$ half-plane, which we can reject as unphysical, and is thus not included in the figure. Of more concern is the prediction that the vacuum lobe and second Mott lobe are overlapping, evident in the overlap of the white lines in Fig. \ref{jch3_betaequals1point5_psi_and_n}(a) in the region $\widetilde{\mu}'\in\lbrace -1.06,1\rbrace,\widetilde{\kappa}\in\lbrace 0,0.35\rbrace$, which is also clearly unphysical but not representative of a discardable solution. A plot of the self-consistently determined value of $\langle\hat{n}\rangle$, in Fig. \ref{jch3_betaequals1point5_psi_and_n}(b), seems to resolve this question with a prediction that the zeroth and second Mott lobes do not overlap, but abruptly meet each other with a boundary at $\widetilde{\kappa}\geq 0$.

\section*{Discussion}
The discrepancy between the numeric and analytical predictions, for $\widetilde{\beta} > \sqrt{2}$, is quite unlike what one observes in the Bose-Hubbard system \cite{prb_40_1_546-570_1989,pra_63_5_053601_2001}, or the JCH system \cite{pra_80_2_023811_2009}, or even the system in consideration for $\widetilde{\beta}\leq\sqrt{2}$. In these systems it was always assumed that the phase transition is second-order, that is, the order parameter $\psi$ is continuous across the transition. A discontinuous, \textit{first-order} phase transition features a jump in $\psi$ from zero in the disordered phase to a distinct non-zero value in the ordered phase, and the condition $a_2 = 0$ does not describe the value of the phase boundary for such transitions \cite{repprogphys_50_7_783-859_1987}. 

To understand the nature of the transition  it is instructive to recall from Fig. \ref{jch3_betaequals1point5_psi_and_n}(b) that the Mott lobe adjacent to the vacuum corresponds to $\langle\hat{n}\rangle = 2$. The complete absence of the $n = 1$ Mott lobe for $\widetilde{\beta} \geq \sqrt{2}$ must be attributable to the fact that in the two-body attractive regime, there is no value of $\widetilde{\mu}'$ in the limit of zero hopping at which it is energetically favourable for there to be a single polariton at a given site, rather than zero or two of them. However, the existence of the finite-length boundary, as a function of $\widetilde{\kappa}$ (see Fig. \ref{jch3_betaequals1point5_psi_and_n}(b)), between these lobes raises the question of what phase exists between them. In general, two Mott lobes meet at the value of $\widetilde{\mu}'$ such that the two corresponding lowest polariton branches are degenerate. A reasonable expectation would therefore be that the quantity
\begin{equation}
\varphi = \langle\hat{a}_i\hat{a}_i\rangle\,\forall\,i
\end{equation}
is nonzero along this boundary. This corresponds to the mean-field approximation of the order parameter for the \textit{pair superfluid} phase, in which superfluidity occurs due to the condensation of (quasi-)particle pairs\cite{pra_92_4_043615_2015}. The true pair superfluid phase occurs when $\psi = 0$ and $\varphi \neq 0$, and is associated with the partial breaking of the system's $\mathbb{U}(1)$ symmetry to $\mathbb{U}(1)/\left(\mathbb{Z}/2\mathbb{Z}\right)$\cite{prl_106_18_183302_2011}.

This anomalous behaviour in the phase transitions of the three-level JCH system when $\widetilde{\beta} > \sqrt{2}$ is not entirely unexpected. While the precise details are different, similar behaviour has been noted in the Bose-Hubbard model with attractive onsite two-body interactions and an additional onsite three-body repulsion\cite{prl_92_5_050402_2004}. The relevant Hamiltonian, scaled appropriately such that the interaction strengths are dimensionless, is provided by
\begin{eqnarray}
\widehat{H}^{BH3} = \sum_i\left\lbrace\frac{-\hat{n}_i\left(\hat{n}_i-1\right)}{2} + \frac{\widetilde{W}\hat{n}_i\left(\hat{n}_i-1\right)\left(\hat{n}_i-2\right)}{6} - \widetilde{\mu}\hat{n}_i\right\rbrace -\widetilde{\kappa}\sum_{\left\langle i,j\right\rangle}\hat{a}^{\dagger}_i\hat{a}_j \label{eq:Bose_Hubbard}
\end{eqnarray}
where $\widetilde{\kappa},\widetilde{\mu}$ and $\widetilde{W}$ are all non-negative.

This model has been studied in the context of experimental proposals which suggest that accounting for the three-body loss rate in a Bose-Hubbard model with attractive onsite two-body interactions may potentially result in either an effective finite three-body repulsion mechanism\cite{pra_89_5_053619_2014} or a hardcore three-body constraint\cite{prl_102_4_040402_2009}. We note that in the Hamiltonian described by Eq.~(\ref{eq:Bose_Hubbard}), a hardcore three-body constraint may be modelled by taking the limit $\widetilde{W}\rightarrow\infty$. In the case where $\widetilde{W}$ is finite, a zero-temperature phase diagram has previously been obtained for the one-dimensional chain using exact diagonalisation and the Density Matrix Renormalisation Group (DMRG)\cite{pra_92_4_043615_2015}. Furthermore, in the hardcore limit ($W\rightarrow\infty$), a sophisticated procedure of spin-model mapping followed by the derivation of an effective field theory has provided a rich understanding of both the phase diagram and the dynamics of the system in each phase\cite{prb_82_6_064509_2010,prb_82_6_064510_2010}. These studies demonstrate that in the presence of an additional three-body repulsion, the two-body-attractive Bose-Hubbard system exhibits both superfluid ($\psi \neq 0, \varphi \neq 0$) and pair superfluid ($\psi = 0, \varphi \neq 0$) phases, with the phase transition between the pair superfluid and the Mott phase being first-order \cite{prl_102_4_040402_2009,prb_82_6_064509_2010,prb_82_6_064510_2010,pra_81_6_061604_2010,prl_106_18_183302_2011,pra_92_4_043615_2015}. Noting the similarities between this system and the three-level JCH system studied here, we predict that when $\widetilde{\beta} > \sqrt{2}$, the three-level JCH system exhibits the pair superfluid phase and an associated pair superfluid-Mott insulator first-order phase transition.

However, it is also noted that in these previous studies of similar systems, the phase boundary between the zeroth and second Mott phases is resolved in such a manner that the pair superfluid phase is not merely confined to a straight line at $\widetilde{\mu} = -0.5$ but in a narrow but nonetheless finite band of parameter space. It is not possible to explore this feature rigorously via the mean-field formalism employed in this work due to the somewhat crude approximation, made \textit{vis-a-vis} spatially off-diagonal correlations, inherent in the mean-field approximation. It would therefore seem that utilising one or more of the methods employed for the three-body-repulsive, two-body-attractive Bose-Hubbard model -- which include an effective potential mechanism with both single and paired source fields\cite{pra_81_6_061604_2010}, and quantum Monte Carlo methods\cite{prl_106_18_183302_2011} -- would serve to elucidate the possible pair superfluid phase in the three-level JCH system.

The addition of a third energy level to each `atom' in a Jaynes-Cummings cavity array is seen here to result in an additional effective three-body repulsion between the resonant cavity photons, analogous to existing modified Bose-Hubbard models with an additional onsite three-body repulsion. By uniformly adjusting the ratio of the two transition dipole moments of each atom, we predict that a crossover manifests itself, between the regimes of two-body repulsion and attraction. Strikingly, this is seen to result in a possible first-order phase transition between the superfluid and Mott vacuum phases, and the coexistence of first- and second-order superfluid-Mott insulator phase transitions across the second Mott lobe. Furthermore, the first Mott lobe vanishes in the regime of two-body attraction.

However, the presence of a pair superfluid phase, which our results seem to suggest, is not provable via the methods employed here. Its existence in the three-body constrained Bose-Hubbard model, demonstrated via exact diagonalisation, effective potential and quantum Monte Carlo simulations, motivated our hypothesis that it is also present in this modified Jaynes-Cummings-Hubbard system. Thus it would seem prudent to employ one or more of these to verify the existence of a pair superfluid of polaritons in such a coupled-cavity array.

\section*{Methods}
\subsection*{Two-Body Repulsion-Attraction Crossover at $\widetilde{\beta}^2 = 2$}
Below we show that  the crossover from two-body repulsion to two-body attraction in the modified cavity system occurs when $\widetilde{\beta}^2 = 2$, a result independent of $\widetilde{\Delta}$. This may be seen by equating $E_1'$ and $E_2'$, i.e.
\begin{equation}
\frac{1}{2}\left\lbrace -2\widetilde{\mu}' - \widetilde{\Delta} + \xi_1^{(2)}\right\rbrace = -\widetilde{\mu}' -\frac{\widetilde{\Delta}}{2} - \sqrt{1+\frac{\widetilde{\Delta}^2}{4}}.
\end{equation}
From Eqs.~(\ref{eq:16},\ref{eq:arcosine},\ref{eq:coeffs_xi}), this is equivalent to
\begin{equation}
\sqrt{\frac{p_2}{3}}\cos\left[\frac{\arccos\left(\frac{4\widetilde{\Delta}(2-\widetilde{\beta}^2)}{u_2^3}\right)+2\pi}{3}\right] = -\sqrt{1+\frac{\widetilde{\Delta}^2}{4}}. \label{eq:crossover}
\end{equation}
The left-hand side of Eq.~(\ref{eq:crossover}) is equal to $-\sqrt{p_2}/2$ when $\widetilde{\beta}^2 = 2$, and recalling from Eq.~({\ref{eq:coeffs_xi}) that $p_2(\widetilde{\beta}^2 = 2) = \widetilde{\Delta}^2 + 4$,
the equality $E_1'(\widetilde{\beta}^2) = E_2'(\widetilde{\beta}^2)$ follows when $\widetilde{\beta}^2=2$.

\subsection*{Analytical Expressions for the Phase Boundaries In The Three-Level JCH System}
From Eqs.~(\ref{eq:En0},\ref{eq:En1},\ref{eq:En2},\ref{eq:landau_0}) we have
\begin{eqnarray}
a_0(n = 0) = 0, \,\,\, 
a_0(n = 1) = -\widetilde{\mu}' - \frac{\widetilde{\Delta}}{2} - \sqrt{1 + \frac{\widetilde{\Delta}^2}{4}} \,\,\, {\rm and} \,\,\,
a_0(n\geq 2) = -n\tilde{\mu}' - \widetilde{\Delta} + \xi_1^{(n)}. \nonumber
\end{eqnarray}
While $a_0$ is not required for the determination of the phase boundaries at second-order phase transitions, we include them for the sake of completeness. However, we do require the second-order Landau coefficients, $a_2' = a_2 - \widetilde{\kappa}\psi^2$, which are given by
\begin{eqnarray}
a_2'(n = 0) &=& \frac{\left|\langle 0|\widehat{V}| 1\rangle_{+}\right|^2}{E(0) - E_{+}(1)} + \frac{\left| |0\rangle |\hat{V}|1\rangle_{-}\right|^2}{E(0) - E_{-}(1)} \\
a_2'(n = 1) &=& \frac{\left|\langle 1|_{-}\widehat{V}|0\rangle\right|^2}{E_{-}(1) - E(0)} + \sum_{k = 1}^3\frac{\left|\langle1|_{-}\widehat{V}|2\rangle_k\right|^2}{E_{-}(1) - E_k(2)} \\
a_2'(n\geq 2) &=& \sum_{k = 1}^3\left\lbrace\frac{\left|\langle n|_{1}\widehat{V}|n-1\rangle_k\right|^2}{E_1(n) - E_k(n-1)} \right. + \left.\frac{\left|\langle n|_{1}\widehat{V}|n+1\rangle_k\right|^2}{E_1(n) - E_k(n+1)}\right\rbrace
\end{eqnarray}
where $E_{k}(n)$ represents the unperturbed energy eigenvalues. We note here that a solution to $a_2(n) = 0$, for a given value of $\widetilde{\mu}'$, may only be utilised to describe the superfluid-Mott insulator phase boundary if, at $\widetilde{\kappa} = 0$, the ground state is given by the lowest polariton branch corresponding to $\langle\hat{n}\rangle = n$. Unphysical solutions such as Mott lobes seemingly extending to negative values of $\widetilde{\kappa}$ can otherwise arise; these are necessarily discarded. 


\begin{thebibliography}{10}
\expandafter\ifx\csname url\endcsname\relax
  \def\url#1{\texttt{#1}}\fi
\expandafter\ifx\csname urlprefix\endcsname\relax\def\urlprefix{URL }\fi
\expandafter\ifx\csname doiprefix\endcsname\relax\def\doiprefix{DOI }\fi
\providecommand{\bibinfo}[2]{#2}
\providecommand{\eprint}[2][]{\url{#2}}

\bibitem{natphys_2_12_849-855_2006}
\bibinfo{author}{Hartmann, M.~J.}, \bibinfo{author}{Brandao, F. G. S.~L.} \&
  \bibinfo{author}{Plenio, M.~B.}
\newblock \bibinfo{journal}{\bibinfo{title}{Strongly interacting polaritons in
  coupled arrays of cavities}}.
\newblock {\emph{\JournalTitle{Nature Physics}}} \textbf{\bibinfo{volume}{2}},
  \bibinfo{pages}{849} (\bibinfo{year}{2006}).

\bibitem{natphys_2_12_856-861_2006}
\bibinfo{author}{Greentree, A.~D.}, \bibinfo{author}{Tahan, C.},
  \bibinfo{author}{Cole, J.~H.} \& \bibinfo{author}{Hollenberg, L. C.~L.}
\newblock \bibinfo{journal}{\bibinfo{title}{Quantum phase transitions of
  light}}.
\newblock {\emph{\JournalTitle{Nature Physics}}} \textbf{\bibinfo{volume}{2}},
  \bibinfo{pages}{856} (\bibinfo{year}{2006}).

\bibitem{pra_76_3_031805_2007}
\bibinfo{author}{Angelakis, D.~G.}, \bibinfo{author}{Santos, M.~F.} \&
  \bibinfo{author}{Bose, S.}
\newblock \bibinfo{journal}{\bibinfo{title}{Photon-blockade-induced {Mott}
  transitions and {XY} spin models in coupled cavity arrays}}.
\newblock {\emph{\JournalTitle{Physical Review A}}}
  \textbf{\bibinfo{volume}{76}}, \bibinfo{pages}{031805(R)}
  (\bibinfo{year}{2007}).

\bibitem{prl_103_8_086403_2009}
\bibinfo{author}{Schmidt, S.} \& \bibinfo{author}{Blatter, G.}
\newblock \bibinfo{journal}{\bibinfo{title}{Strong coupling theory for the
  {Jaynes-Cummings-Hubbard} model}}.
\newblock {\emph{\JournalTitle{Physical Review Letters}}}
  \textbf{\bibinfo{volume}{103}}, \bibinfo{pages}{086403}
  (\bibinfo{year}{2009}).

\bibitem{pra_80_2_023811_2009}
\bibinfo{author}{Koch, J.} \& \bibinfo{author}{Hur, K.~L.}
\newblock \bibinfo{journal}{\bibinfo{title}{Superfluid-{Mott}-insulator
  transition of light in the {Jaynes-Cummings} lattice}}.
\newblock {\emph{\JournalTitle{Physical Review A}}}
  \textbf{\bibinfo{volume}{80}}, \bibinfo{pages}{023811}
  (\bibinfo{year}{2009}).

\bibitem{prl_104_21_216402_2010}
\bibinfo{author}{Schmidt, S.} \& \bibinfo{author}{Blatter, G.}
\newblock \bibinfo{journal}{\bibinfo{title}{Excitations of strongly correlated
  lattice polaritons}}.
\newblock {\emph{\JournalTitle{Physical Review Letters}}}
  \textbf{\bibinfo{volume}{104}}, \bibinfo{pages}{216402}
  (\bibinfo{year}{2010}).

\bibitem{pra_81_6_061801r_2010}
\bibinfo{author}{Tomadin, A.} \emph{et~al.}
\newblock \bibinfo{journal}{\bibinfo{title}{Signatures of the
  superfluid-insulator phase transition in laser-driven dissipative nonlinear
  cavity arrays}}.
\newblock {\emph{\JournalTitle{Physical Review A}}}
  \textbf{\bibinfo{volume}{81}}, \bibinfo{pages}{061801(R)}
  (\bibinfo{year}{2010}).

\bibitem{pra_90_4_043801_2014}
\bibinfo{author}{Bujnowski, B.}, \bibinfo{author}{Corso, J.~K.},
  \bibinfo{author}{Hayward, A. L.~C.}, \bibinfo{author}{Cole, J.~H.} \&
  \bibinfo{author}{Martin, A.~M.}
\newblock \bibinfo{journal}{\bibinfo{title}{Supersolid phases of light in
  extended {Jaynes-Cummings-Hubbard} systems}}.
\newblock {\emph{\JournalTitle{Physical Review A}}}
  \textbf{\bibinfo{volume}{90}}, \bibinfo{pages}{043801}
  (\bibinfo{year}{2014}).

\bibitem{prl_99_18_186401_2007}
\bibinfo{author}{Rossini, D.} \& \bibinfo{author}{Fazio, R.}
\newblock \bibinfo{journal}{\bibinfo{title}{Mott-insulating and glassy phases
  of polaritons in 1{D} arrays of coupled cavities}}.
\newblock {\emph{\JournalTitle{Physical Review Letters}}}
  \textbf{\bibinfo{volume}{99}}, \bibinfo{pages}{186401}
  (\bibinfo{year}{2007}).

\bibitem{optexpress_19_12_11018-11033_2011}
\bibinfo{author}{Quach, J.~Q.}, \bibinfo{author}{Su, C.-H.},
  \bibinfo{author}{Martin, A.~M.}, \bibinfo{author}{Greentree, A.~D.} \&
  \bibinfo{author}{Hollenberg, L. C.~L.}
\newblock \bibinfo{journal}{\bibinfo{title}{Reconfigurable quantum
  metamaterials}}.
\newblock {\emph{\JournalTitle{Optics Express}}} \textbf{\bibinfo{volume}{19}},
  \bibinfo{pages}{11018} (\bibinfo{year}{2011}).

\bibitem{prl_101_24_246809_2008}
\bibinfo{author}{Cho, J.}, \bibinfo{author}{Angelakis, D.~G.} \&
  \bibinfo{author}{Bose, S.}
\newblock \bibinfo{journal}{\bibinfo{title}{Fractional quantum {Hall} state in
  coupled cavities}}.
\newblock {\emph{\JournalTitle{Physical Review Letters}}}
  \textbf{\bibinfo{volume}{101}}, \bibinfo{pages}{246809}
  (\bibinfo{year}{2008}).

\bibitem{prl_108_20_206809_2012}
\bibinfo{author}{Umucalilar, R.~O.} \& \bibinfo{author}{Carusotto, I.}
\newblock \bibinfo{journal}{\bibinfo{title}{Fractional quantum {Hall} states of
  photons in an array of dissipative coupled cavities}}.
\newblock {\emph{\JournalTitle{Physical Review Letters}}}
  \textbf{\bibinfo{volume}{108}}, \bibinfo{pages}{206809}
  (\bibinfo{year}{2012}).

\bibitem{prl_108_22_223602_2012}
\bibinfo{author}{Hayward, A. L.~C.}, \bibinfo{author}{Martin, A.~M.} \&
  \bibinfo{author}{Greentree, A.~D.}
\newblock \bibinfo{journal}{\bibinfo{title}{Fractional quantum {Hall} physics
  in {Jaynes-Cummings-Hubbard} lattices}}.
\newblock {\emph{\JournalTitle{Physical Review Letters}}}
  \textbf{\bibinfo{volume}{108}}, \bibinfo{pages}{223602}
  (\bibinfo{year}{2012}).

\bibitem{pra_93_5_053614_2016}
\bibinfo{author}{Hayward, A. L.~C.} \& \bibinfo{author}{Martin, A.~M.}
\newblock \bibinfo{journal}{\bibinfo{title}{Pfaffian states in coupled
  atom-cavity systems}}.
\newblock {\emph{\JournalTitle{Physical Review A}}}
  \textbf{\bibinfo{volume}{93}}, \bibinfo{pages}{053614}
  (\bibinfo{year}{2016}).

\bibitem{nuclphysb_360_2-3_362-396_1991}
\bibinfo{author}{Moore, G.} \& \bibinfo{author}{Read, N.}
\newblock \bibinfo{journal}{\bibinfo{title}{Nonabelions in the fractional
  quantum hall effect}}.
\newblock {\emph{\JournalTitle{Nuclear Physics B}}}
  \textbf{\bibinfo{volume}{360}}, \bibinfo{pages}{362} (\bibinfo{year}{1991}).

\bibitem{natphys_3_10_726-731_2007}
\bibinfo{author}{B\"{u}chler, H.~P.}, \bibinfo{author}{Micheli, A.} \&
  \bibinfo{author}{Zoller, P.}
\newblock \bibinfo{journal}{\bibinfo{title}{Three-body interactions with cold
  polar molecules}}.
\newblock {\emph{\JournalTitle{Nature Physics}}} \textbf{\bibinfo{volume}{3}},
  \bibinfo{pages}{726} (\bibinfo{year}{2007}).

\bibitem{prl_102_4_040402_2009}
\bibinfo{author}{Daley, A.~J.}, \bibinfo{author}{Taylor, J.~M.},
  \bibinfo{author}{Diehl, S.}, \bibinfo{author}{Baranov, M.} \&
  \bibinfo{author}{Zoller, P.}
\newblock \bibinfo{journal}{\bibinfo{title}{Atomic three-body loss as a
  dynamical three-body interaction}}.
\newblock {\emph{\JournalTitle{Physical Review Letters}}}
  \textbf{\bibinfo{volume}{102}}, \bibinfo{pages}{040402}
  (\bibinfo{year}{2009}).

\bibitem{pra_82_4_043629_2010}
\bibinfo{author}{Mazza, L.}, \bibinfo{author}{Rizzi, M.},
  \bibinfo{author}{Lewenstein, M.} \& \bibinfo{author}{Cirac, J.~I.}
\newblock \bibinfo{journal}{\bibinfo{title}{Emerging bosons with three-body
  interactions from spin-1 atoms in optical lattices}}.
\newblock {\emph{\JournalTitle{Physical Review A}}}
  \textbf{\bibinfo{volume}{82}}, \bibinfo{pages}{043629}
  (\bibinfo{year}{2010}).

\bibitem{pra_89_5_053619_2014}
\bibinfo{author}{Daley, A.~J.} \& \bibinfo{author}{Simon, J.}
\newblock \bibinfo{journal}{\bibinfo{title}{Effective three-body interactions
  via photon-assisted tunneling in an optical lattice}}.
\newblock {\emph{\JournalTitle{Physical Review A}}}
  \textbf{\bibinfo{volume}{89}}, \bibinfo{pages}{053619}
  (\bibinfo{year}{2014}).

\bibitem{prb_40_1_546-570_1989}
\bibinfo{author}{Fisher, M. P.~A.}, \bibinfo{author}{Weichman, P.~B.},
  \bibinfo{author}{Grinstein, G.} \& \bibinfo{author}{Fisher, D.~S.}
\newblock \bibinfo{journal}{\bibinfo{title}{Boson localization and the
  superfluid-insulator transition}}.
\newblock {\emph{\JournalTitle{Physical Review B}}}
  \textbf{\bibinfo{volume}{40}}, \bibinfo{pages}{546} (\bibinfo{year}{1989}).

\bibitem{prb_82_6_064509_2010}
\bibinfo{author}{Diehl, S.}, \bibinfo{author}{Baranov, M.},
  \bibinfo{author}{Daley, A.~J.} \& \bibinfo{author}{Zoller, P.}
\newblock \bibinfo{journal}{\bibinfo{title}{Quantum field theory for the
  three-body constrained lattice {Bose} gas. {I. Formal} developments}}.
\newblock {\emph{\JournalTitle{Physical Review B}}}
  \textbf{\bibinfo{volume}{82}}, \bibinfo{pages}{064509}
  (\bibinfo{year}{2010}).

\bibitem{prb_82_6_064510_2010}
\bibinfo{author}{Diehl, S.}, \bibinfo{author}{Baranov, M.},
  \bibinfo{author}{Daley, A.~J.} \& \bibinfo{author}{Zoller, P.}
\newblock \bibinfo{journal}{\bibinfo{title}{Quantum field theory for the
  three-body constrained lattice {Bose} gas. {II. Application} to the many-body
  problem}}.
\newblock {\emph{\JournalTitle{Physical Review B}}}
  \textbf{\bibinfo{volume}{82}}, \bibinfo{pages}{064510}
  (\bibinfo{year}{2010}).

\bibitem{pra_81_6_061604_2010}
\bibinfo{author}{Lee, Y.-W.} \& \bibinfo{author}{Yang, M.-F.}
\newblock \bibinfo{journal}{\bibinfo{title}{Superfluid-insulator transitions in
  attractive {Bose-Hubbard} model with three-body constraint}}.
\newblock {\emph{\JournalTitle{Physical Review A}}}
  \textbf{\bibinfo{volume}{81}}, \bibinfo{pages}{061604(R)}
  (\bibinfo{year}{2010}).

\bibitem{prl_106_18_183302_2011}
\bibinfo{author}{Bonnes, L.} \& \bibinfo{author}{Wessel, S.}
\newblock \bibinfo{journal}{\bibinfo{title}{Pair superfluidity of three-body
  constrained bosons in two dimensions}}.
\newblock {\emph{\JournalTitle{Physical Review Letters}}}
  \textbf{\bibinfo{volume}{106}}, \bibinfo{pages}{185302}
  (\bibinfo{year}{2011}).

\bibitem{pra_92_4_043615_2015}
\bibinfo{author}{Sowi\'{n}ski, T.}, \bibinfo{author}{Chhajlany, R.~W.},
  \bibinfo{author}{Dutta, O.}, \bibinfo{author}{Tagliacozzo, L.} \&
  \bibinfo{author}{Lewenstein, M.}
\newblock \bibinfo{journal}{\bibinfo{title}{Criticality in the {Bose-Hubbard}
  model with three-body repulsion}}.
\newblock {\emph{\JournalTitle{Physical Review A}}}
  \textbf{\bibinfo{volume}{92}}, \bibinfo{pages}{043615}
  (\bibinfo{year}{2015}).

\bibitem{procieee_51_1_89-109_1963}
\bibinfo{author}{Jaynes, E.~T.} \& \bibinfo{author}{Cummings, F.~W.}
\newblock \bibinfo{journal}{\bibinfo{title}{Comparison of quantum and
  semiclassical radiation theories with application to the beam maser}}.
\newblock {\emph{\JournalTitle{Proceedings of the IEEE}}}
  \textbf{\bibinfo{volume}{51}}, \bibinfo{pages}{89} (\bibinfo{year}{1963}).

\bibitem{klimovchumakovqtmoptics}
\bibinfo{author}{Klimov, A.~B.} \& \bibinfo{author}{Chumakov, S.~M.}
\newblock \emph{\bibinfo{title}{A Group-Theoretical Approach to Quantum Optics:
  Models of Atom-Field Interactions}} (\bibinfo{publisher}{Wiley-VCH},
  \bibinfo{year}{2009}).

\bibitem{atomicphysicsfoot}
\bibinfo{author}{Foot, C.~J.}
\newblock \emph{\bibinfo{title}{Atomic Physics}} (\bibinfo{publisher}{Oxford
  University Press}, \bibinfo{year}{2005}).

\bibitem{mathmethphysarfkenweber}
\bibinfo{author}{Arfken, G.~B.}, \bibinfo{author}{Weber, H.~J.} \&
  \bibinfo{author}{Harris, F.~E.}
\newblock \emph{\bibinfo{title}{Mathematical Methods for Physicists}}
  (\bibinfo{publisher}{Academic Press}, \bibinfo{year}{2012}),
  \bibinfo{edition}{7} edn.

\bibitem{mathgaz_77_480_354-359_1993}
\bibinfo{author}{Nickalls, R. W.~D.}
\newblock \bibinfo{journal}{\bibinfo{title}{A new approach to solving the
  cubic: {Cardan's} solution revealed}}.
\newblock {\emph{\JournalTitle{Mathematical Gazette}}}
  \textbf{\bibinfo{volume}{77}}, \bibinfo{pages}{354} (\bibinfo{year}{1993}).

\bibitem{physscr_76_3_244-248_2007}
\bibinfo{author}{Abdel-Wahab, N.~H.}
\newblock \bibinfo{journal}{\bibinfo{title}{A three-level atom interacting with
  a single mode cavity field: Different configurations}}.
\newblock {\emph{\JournalTitle{Physica Scripta}}}
  \textbf{\bibinfo{volume}{76}}, \bibinfo{pages}{244} (\bibinfo{year}{2007}).

\bibitem{pra_46_11_r6801r_1992}
\bibinfo{author}{Tian, L.} \& \bibinfo{author}{Carmichael, H.~J.}
\newblock \bibinfo{journal}{\bibinfo{title}{Quantum trajectory simulations of
  the two-state behavior of an optical cavity containing one atom}}.
\newblock {\emph{\JournalTitle{Physical Review A}}}
  \textbf{\bibinfo{volume}{46}}, \bibinfo{pages}{R6801(R)}
  (\bibinfo{year}{1992}).

\bibitem{nature_436_7047_87-90_2005}
\bibinfo{author}{Birnbaum, K.~M.} \emph{et~al.}
\newblock \bibinfo{journal}{\bibinfo{title}{Photon blockade in an optical
  cavity with one trapped atom}}.
\newblock {\emph{\JournalTitle{Nature}}} \textbf{\bibinfo{volume}{436}},
  \bibinfo{pages}{87} (\bibinfo{year}{2005}).

\bibitem{pra_63_5_053601_2001}
\bibinfo{author}{van Oosten, D.}, \bibinfo{author}{van~der Straten, P.} \&
  \bibinfo{author}{Stoof, H. T.~C.}
\newblock \bibinfo{journal}{\bibinfo{title}{Quantum phases in an optical
  lattice}}.
\newblock {\emph{\JournalTitle{Physical Review A}}}
  \textbf{\bibinfo{volume}{63}}, \bibinfo{pages}{053601}
  (\bibinfo{year}{2001}).

\bibitem{sachdevquantumpt}
\bibinfo{author}{Sachdev, S.}
\newblock \emph{\bibinfo{title}{Quantum Phase Transitions}}
  (\bibinfo{publisher}{Cambridge University Press}, \bibinfo{year}{2011}),
  \bibinfo{edition}{2} edn.

\bibitem{sakuraimodernqm}
\bibinfo{author}{Sakurai, J.~J.} \& \bibinfo{author}{Napolitano, J.~J.}
\newblock \emph{\bibinfo{title}{Modern Quantum Mechanics}}
  (\bibinfo{publisher}{Addison-Wesley}, \bibinfo{year}{2011}),
  \bibinfo{edition}{2} edn.

\bibitem{repprogphys_50_7_783-859_1987}
\bibinfo{author}{Binder, K.}
\newblock \bibinfo{journal}{\bibinfo{title}{Theory of first-order phase
  transitions}}.
\newblock {\emph{\JournalTitle{Reports on Progress in Physics}}}
  \textbf{\bibinfo{volume}{50}}, \bibinfo{pages}{783} (\bibinfo{year}{1987}).

\bibitem{prl_92_5_050402_2004}
\bibinfo{author}{Kuklov, A.}, \bibinfo{author}{Prokof'ev, N.~V.} \&
  \bibinfo{author}{Svistunov, B.}
\newblock \bibinfo{journal}{\bibinfo{title}{Commensurate two-component bosons
  in an optical lattice: Ground state phase diagram}}.
\newblock {\emph{\JournalTitle{Physical Review Letters}}}
  \textbf{\bibinfo{volume}{92}}, \bibinfo{pages}{050402}
  (\bibinfo{year}{2004}).

\end{thebibliography}

\section*{Acknowledgements}
The authors would to thank A. L. C. Hayward for fruitful discussions regarding the Jaynes-Cummings-Hubbard model. S. B. P. was supported by an Australian Government Research Training Program Scholarship and by the University of Melbourne. A. M. M. would like to thank the Institute of Advanced Study (Durham University, U.K.) for hosting him during the preparation of this manuscript.

\section*{Author contributions statement}
A.A.M. conceived the calculations, S.B.P. conducted the calculations, A.M.M. and S.B.P. jointly interpreted the results and co-authored the manuscript. 

\section*{Additional Information}
\subsection*{Competing interests}
The authors declare no competing interests.

\end{document}